\documentclass{article}

\usepackage{arxiv}

\usepackage[utf8]{inputenc} % allow utf-8 input
\usepackage[T1]{fontenc}    % use 8-bit T1 fonts
\usepackage{hyperref}       % hyperlinks
\usepackage{url}            % simple URL typesetting
\usepackage{booktabs}       % professional-quality tables
\usepackage{amsfonts}       % blackboard math symbols
\usepackage{nicefrac}       % compact symbols for 1/2, etc.
\usepackage{microtype}      % microtypography
\usepackage{lipsum}		% Can be removed after putting your text content
\usepackage{graphicx}
\usepackage{natbib}
\usepackage{doi}

\usepackage{xcolor}

\newcommand{\figref}[1]{Figure~\ref{fig:#1}}
\newcommand{\secref}[1]{Section~\ref{sec:#1}}

\LetLtxMacro{\originaleqref}{\eqref}
\newcommand{\eqref}[1]{Eq.~\ref{eq:#1}}

\title{HDR-VDP-3: A multi-metric for predicting image differences, quality and contrast distortions in high dynamic range and regular content}

\date{} 					% Or removing it

\author{ \href{https://orcid.org/0000-0003-2353-0349}{\includegraphics[scale=0.06]{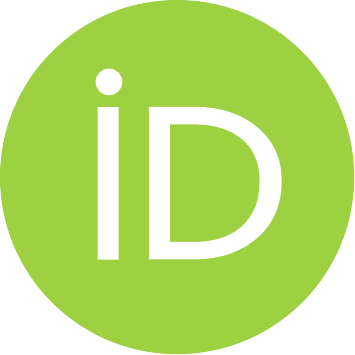}\hspace{1mm}Rafa{\l} K. Mantiuk}\\
	University of Cambridge\\
	\texttt{rafal.mantiuk@cl.cam.ac.uk} \\
	\And
    \href{https://orcid.org/0000-0001-7475-6722}{\includegraphics[scale=0.06]{orcid.pdf}\hspace{1mm}Dounia Hammou} \\
	University of Cambridge\\
	\texttt{dh706@cl.cam.ac.uk} \\
    \And
    \href{https://orcid.org/0000-0002-7985-4177}{\includegraphics[scale=0.06]{orcid.pdf}\hspace{1mm}Param Hanji} \\
	University of Cambridge\\
	\texttt{pmh64@cl.cam.ac.uk} \\
}

\renewcommand{\headeright}{}
\renewcommand{\undertitle}{}
\renewcommand{\shorttitle}{}

\hypersetup{
pdftitle={HDR-VDP-3: A multi-metric for predicting image differences, quality and contrast distortions in high dynamic range and standard content},
pdfsubject={HDR-VDP-3},
pdfauthor={Rafa{\l} K. Mantiuk, Dounia Hammou, Param Hanji},
pdfkeywords={Image Metric, High Dynamic Range, Human Vision, Display-adaptive quality},
}

\begin{document}
\maketitle

\begin{abstract}
High-Dynamic-Range Visual-Difference-Predictor version 3, or HDR-VDP-3, is a visual metric that can fulfill several tasks, such as full-reference image/video quality assessment, prediction of visual differences between a pair of images, or prediction of contrast distortions. Here we present a high-level overview of the metric, position it with respect to related work, explain the main differences compared to version 2.2, and describe how the metric was adapted for the \href{https://sites.google.com/view/wacv2023-workshop-quality-va/competition}{HDR Video Quality Measurement Grand Challenge 2023}.
\end{abstract}

% keywords can be removed
\keywords{Image Metric \and High Dynamic Range}

\section{Introduction}

High-Dynamic-Range Visual-Difference-Predictor version 3, or HDR-VDP-3 is an image metric that can address multiple applications: the prediction of image quality, visible differences, and contrast distortions. If we want to optimize for image quality, for example, by selecting the right resolution and compression configuration for video streaming, we want to use \emph{a full-reference quality metric} that compares a distorted image/video (e.g., decoded frame) to its reference and assesses the overall magnitude of introduced distortion. However, if we want to ensure that the introduced distortions are invisible, for example, for visually lossless compression, we may want to use \emph{a visibility metric}, which predicts the probability of detecting differences in each part of the image. In other cases, we may want to test an image processing algorithm that modifies an image but should not introduce disturbing artifacts. An example is tone mapping, where an image tone mapped for a low-dynamic range display must be different from the input high-dynamic-range image, but it should preserve the general visibility of the contrast. For such applications, we want to use \emph{a contrast distortion metric}. HDR-VDP-3 addresses all three applications using the same core visual model but uses different parameters and final processing stages to provide predictions for each application. 

This short paper is not meant to be a complete description of the metric but a rather high-level overview with references to the relevant papers, which provide further details. HDR-VDP-3 has the same processing pipeline as HDR-VDP-2, which is explained in detail in \citep{Mantiuk2011a}. In this short paper, we first position HDR-VDP-3 with respect to other metrics (\secref{related}), explain the list of tasks it can perform (\secref{hdr-vdp-tasks}), give a high-level overview (\secref{overview}), itemize the differences with respect to HDR-VDP-2 (\secref{changes}) and finally explain how the metric was adapted to perform quality assessment for the WACV HDR Video Quality Measurement Grand Challenge (\secref{challenge}). 

\section{Related metrics}
\label{sec:related}

HDR-VDP-3 is the third major iteration of the metric, which was originally inspired by seminal works on visibility and detection metrics by Daly \citeyear{Daly1993}, Lubin \citeyear{lubin1995visual}, and Watson \citeyear{Watson2000a}. The original HDR-VDP-1 \citep{Mantiuk2005a} was an extension of the VDP by Daly \citeyear{Daly1993}. The extension incorporated changes allowing to compare high dynamic range images: the models of glare, photoreceptor response \citep{Mantiuk2006c} (precursor of the PQ function later used for HDR coding), and contrast sensitivity, which adapts to local luminance. Similar to the VDP, this metric focused on predicting visible difference maps, and it did not provide single-value quality predictions. This was addressed in HDR-VDP-2 \citep{Mantiuk2011a}, which was a major redesign of the original VDP metric: it incorporated separate pathways for rod and cone vision, replaced cortex transform with steerable pyramids (for performance and accuracy), incorporated a new contrast masking model with intra- and inter-channel masking, and provided the predictions of both visual difference maps and image quality. But probably the most significant difference was that HDR-VDP-2 was extensively recalibrated and tested on a large range of basic psychophysical detection and discrimination data. When HDR-VDP-2 was released, its quality predictions could be calibrated only on standard dynamic range image datasets (TID2008 and LIVE) as no HDR quality datasets were available. This was rectified in HDR-VDP-2.2 \citep{Narwaria}, which used two new HDR datasets in addition to TID2008 and CSIQ to recalibrate quality predictions. 

A few important works led to the development of HDR-VDP-3. First, new components were added to simulate the effect of aging on the visual system \citep{Mantiuk2018}: the age-dependent model of glare, crystalline lens aging and senile miosis (reduced pupil dilation in an older eye). Second, a series of new measurements on an HDR display let us model the effect of adaptation to local luminance \citep{Vangorp2015}. Finally, our effort to combine multiple HDR and SDR datasets and bring them to the same quality scale \citep{Perez-Ortiz2018} let us recalibrate the metric on the largest HDR image quality dataset (of over 4000 images) --- UPIQ \citep{Mikhailiuk2021}. Other major changes are discussed in \secref{changes}.

A critical component of the metric is the perceptually uniform encoding of luminance. Such encoding was shown to be an effective method of representing and compressing HDR video \citep{Mantiuk2004b,Mantiuk2006c}, and its refined version was later standardized as a Perceptual Quantizer (SMPTE ST 2084) \citep{Miller2013}. However, we have also demonstrated that such perceptually uniform (PU) encoding can be used to adapt existing SDR quality metrics to HDR images \citep{Aydn2008,Mantiuk2021}.

Parallel to the work on visibility and quality predictions, we also worked on predicting contrast distortions caused by tone-mapping \citep{Aydin2008b}. A modern implementation of this metric is one of the "tasks" of HDR-VDP-3 (\secref{hdr-vdp-tasks}).

Both us \citep{Wolski2018,Ye2019metric} and others \citep{Banterle2020} made an attempt to replace existing HDR-VDP-2 and HDR-VDP-3 metrics with deep-learning architectures. Neural networks bring the advantage of potentially faster processing speeds, no-reference predictions \citep{Banterle2020}, higher accuracy, and easier re-calibration. Although deep-learning metrics show promising results in selected applications, such as visually lossless coding \citep{Ye2019vlc}, they are not explainable and often suffer from over-fitting, as image and quality datasets are typically small in size, and the measurements tend to be noisy. 

More recently, we released Foveated Video VDP (FovVideoVDP) \citep{Mantiuk2021fvvdp}, a metric intended to predict quality in video, assuming a gaze point (foveated viewing) or assuming that the user can look everywhere (as in traditional video quality metrics). FovVideoVDP is a simplified version of HDR-VDP-3, which adds temporal processing (sustained and transient visual channels) and a contrast sensitivity function that accounts for the distance from the gaze location (eccentricity). The new implementation runs on a GPU (both in Matlab and Python/PyTorch) and offers much faster processing speeds. However, because the new metric has not been calibrated on regular (non-foveated) videos and contains multiple simplifications, it provides slightly worse accuracy of predictions. 

\section{HDR-VDP-3 tasks}
\label{sec:hdr-vdp-tasks}

HDR-VDP-3 acts as a predictor of different quantities, depending on the "task" parameter. The tasks include: 

\begin{itemize}
\item [quality] --- the prediction of a single-valued perceived degradation of quality measured in the units of Just-Objectionable-Differences (JOD) \cite[Sec. IVb]{Perez-Ortiz2018}, which are related to the mean-opinion-scores. This task was calibrated using UPIQ dataset \citep{Mikhailiuk2021}. The highest quality is assumed to be 10. The degradation of 1 JOD unit corresponds to 75\% of the population noticing the difference between the pair of images. A further explanation of the JOD scale can be found at \url{https://github.com/gfxdisp/FovVideoVDP#predicted-quality-scores} and in \citep{Perez-Ortiz2018}.
\item [side-by-side] --- the prediction of visible difference maps for pairs of test and reference images presented side-by-side. The values in the map represent the proportion of the population that is likely to notice the difference in a particular part of the image. This task was calibrated using the datasets from \citep{Wolski2018,Ye2019metric}.
\item [flicker] --- similar to the "side-by-side" task, but for test and reference images that are flipped (swapped) every 0.5\,sec. This task was fine-tuned on a smaller dataset, measured in a similar manner as in \citep{Wolski2018} (unpublished).
\item [detection] --- the detection task predicts the probability of detecting the difference between two images (single-valued) and was calibrated on the same datasets as HDR-VDP-2 --- basic psychophysical detection and discrimination data for Gabor patches, sinusoidal gratings, and discs. This task should provide better accuracy for simple stimuli, but potentially lower accuracy for complex images. 
\item [civdm] --- a contrast distortion metric, which is a modern implementation of the dynamic-range-independent visual quality assessment \citep{Aydin2008b}. It predicts the maps which indicate in which image parts the contrast will be lost and in which parts it will be (over-)enhanced. The predictions are different from those of \citep{Aydin2008b} as the original contrast-independent metric was based on HDR-VDP-1.
\end{itemize}

\section{HDR-VDP-3 overview}
\label{sec:overview}

\begin{figure*}
    \centering
    \includegraphics[width=\textwidth]{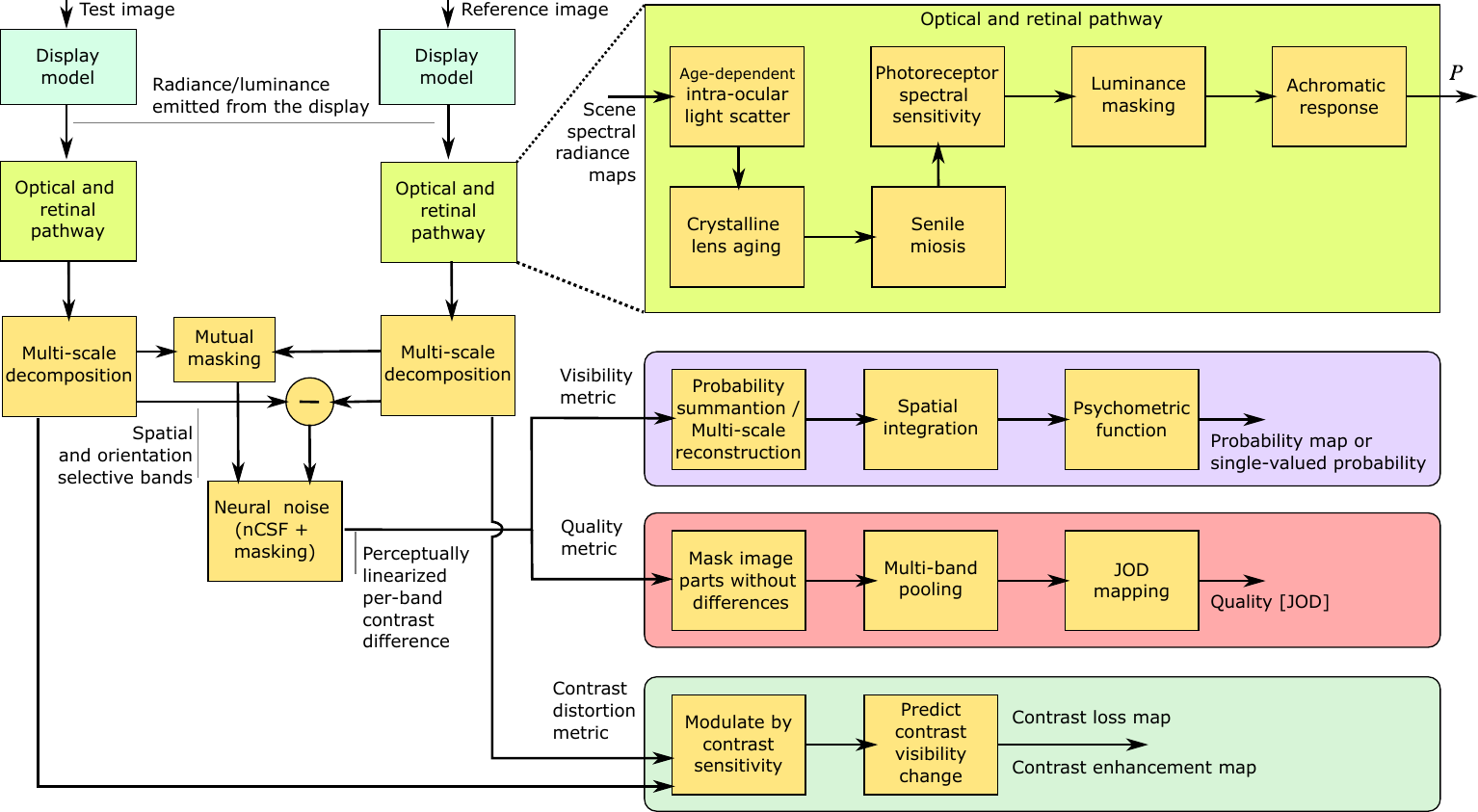}
    \caption{The processing diagram of the HDR-VDP-3. The metric requires the images to be calibrated in the absolute radiance or luminance quantity emitted from a display (HDR or SDR). The physical image representation (radiance map) is processed by the optical and retinal pathway which simulates the eye's optics and photoreceptor responses. The resulting retinal images are then decomposed into multiple scales, each isolating a band of spatial frequencies and orientations. The core component of the metric is the model of contrast masking and (neural) contrast sensitivity, which predicts the visibility of the differences between a pair of images. The multi-band representation from those stages is then fed to one of the different "heads", responsible for the prediction of visibility, quality and contrast distortions. }
    \label{fig:proc-diagram}
\end{figure*}

A high-level overview of the metric is shown in \figref{proc-diagram}. As any full-reference metric, HDR-VDP-3 takes as input a pair of test and reference images. However, those images need to be calibrated in absolute radiometric (or photometric) units by the \emph{display model} since HDR-VDP-3 relies on models of low-level human vision, which operate on photometric units. The \emph{retinal and optical pathway} simulates optics of the eye (glare), age-adaptive lens opacity, pupil, photoreceptor (cones and rods) spectral response, local adaptation and luminance masking. The resulting retinal images are then decomposed into multiple bands of spatial frequencies and orientations. The most important part of the metric is the model of neural contrast sensitivity and contrast masking, which predicts the ability of the visual system to detect and discriminate patterns. The result of that stage is then passed to one of the three "heads" of the metric: one that predicts visibility maps, one that predicts single-values quality and one that predicts contrast distortion maps. The description of those stages is beyond the scope of this short paper, but further details can be found in \citep{Mantiuk2011a,Aydin2008b,Mantiuk2018}.

\section{Differences with respect to HDR-VDP-2}
\label{sec:changes}

Compared to version 2.2 of the metric, HDR-VDP-3 contains the following major changes:
\begin{itemize}
\item It requires specifying a prediction task, as explained in \secref{hdr-vdp-tasks}.
\item It includes a contrast distortion metric, which is a modern implementation of \citep{Aydin2008b}.
\item The contrast sensitivity function was refitted to newer data (from \citep{Joong2013}).
\item The model of glare (MTF) can be disabled or switched to the CIE99 Glare Spread Function \citep{Vos1999}.
\item The metric now accounts for the age-related effects, as described in \citep{Mantiuk2018}.
\item The metric includes a model of local adaptation from \citep{Vangorp2015}.
\item The tasks "side-by-side" and "flicker" have been calibrated on large datasets from \citep{Wolski2018,Ye2019metric}.
\item The task "quality" has been recalibrated using a new UPIQ dataset \citep{Mikhailiuk2021} with over 4000 SDR and HDR images, all scaled in JOD units. 
\item The code now includes multiple examples of how to use the metric in different scenarios. 
\item The code has been reorganised and tested to run on a recent version of Matlab (2022a) but also GNU Octave.
\item The code runs on a GPU (CUDA) in Matlab. 
\end{itemize}

\section{The submission for WACV HDR Video Quality Measurement Grand Challenge}
\label{sec:challenge}

HDR-VDP-3 was submitted to the \href{https://sites.google.com/view/wacv2023-workshop-quality-va/competition}{WACV HDR Video Quality Measurement Grand Challenge 2023}. The organizers of the challenge provided a new HDR video quality dataset --- LIVE HDR \citep{shang2022subjective}.

For the challenge, we used the \emph{quality} task of HDR-VDP-3. We did not consider the temporal aspect of the videos. The metric was applied separately on selected frames, and the scores were averaged to obtain the final video quality score. 
The FFmpeg program \citep{tomar2006converting}  was used to decode every 30th frame in the video and stored as PNG files. Furthermore, a display model consisting of the inverse PQ function \citep{Miller2013} was used to transform the display-encoded pixel values into radiometric (or photometric) units as:
\begin{equation}\label{eq:display_model}
    L(x,y) = PQ^{-1}(I(x,y)) + E_{amb}\frac{k_{refl}}{\pi}\,,
\end{equation}
where $I(x,y)$ is the PQ-encoded RGB frame, $L(x,y)$ is the frame in absolute linear RGB (BT.2020) units (photometric), $E_{amb}$ is the room ambient illumination, and $k_{refl}$ is the display reflectivity. The room ambient illumination and the display reflectivity were set to $200$ and $0.005$ respectively. These terms model the effect of ambient light, which was reported for the experiment.

%To put it together, for each reference and distorted video pair, the HDR-VDP-3 quality metric was used to predict the JOD scores from the selected frames in photometric units. The scores were pooled using the mean function into the video quality JOD score.

It should be mentioned that the HDR-VDP-3 \emph{quality} task has not been (re-)calibrated on the LIVE HDR training dataset. We used the original model, calibrated on the UPIQ dataset. Regardless, the metric was able to correlate well with the mean-opinion scores. 

Although the metric was originally implemented in Matlab, we adapted the code so that it can be run in GNU Octave. The snippet of the code used to run the metric can be found in \verb!examples/hdr_video_pq_eotf.m! in release 3.0.7 of the metric. 

\section*{ACKNOWLEDGMENTS}
This project has received funding from the European Research Council (ERC) under the European Union's Horizon 2020 research and innovation programme (grant agreement N$^\circ$ 725253--EyeCode).

\bibliographystyle{unsrtnat}
\bibliography{references}

\end{document}